\documentclass[preprint,showpacs,preprintnumbers,amsmath,amssymb]{revtex4}
\usepackage{graphicx}% Include figure files
\usepackage{dcolumn}% Align table columns on decimal point
\usepackage{bm}% bold math
\font\scripti=cmmi7
\font\scriptscripti=cmmi5
\def\sib#1{\setbox0 = \hbox{\scripti #1}
  \kern-.02em\copy0\kern-\wd0
  \kern.04em\box0} % script italic bold 
\def\ssib#1{\setbox0 = \hbox{\scriptscripti #1}
  \kern-.02em\copy0\kern-\wd0
  \kern.04em\box0} % scriptscript italic bold
\font\tenib=cmmib10 % italic bold for math
\skewchar\tenib='177 \skewchar\tenib='177 \skewchar\tenib='177
\def\pbold#1{\setbox0 = \hbox{$ #1 $}
  \kern-.022em\copy0\kern-\wd0
  \kern.011em\copy0\kern-\wd0
  \kern.011em\copy0\kern-\wd0
  \kern.011em\copy0\kern-\wd0
  \kern.011em\box0} % poorman's bold
\usepackage{graphicx}% Include figure files
\usepackage{dcolumn}% Align table columns on decimal point
\usepackage{bm}% bold math

\def\up{\uparrow}
\def\dwn{\downarrow}

\def\lesssim{\ \raise.3ex\hbox{$<$}\kern-0.8em\lower.7ex\hbox{$\sim$}\ }
\def\gesim{\ \raise.3ex\hbox{$>$}\kern-0.8em\lower.7ex\hbox{$\sim$}\ }

\begin{document}
\title{Pseudogap phenomenon and effects of population imbalance in the normal state of a unitary Fermi gas}

\author{Takashi Kashimura, Ryota Watanabe, and Yoji Ohashi}
\affiliation{Department of Physics, Keio University, 3-14-1 Hiyoshi, Kohoku-ku, Yokohama 223-8522, Japan} 

\date{\today}
\begin{abstract}
We investigate strong-coupling corrections to single-particle excitations in the normal state of a spin-polarized unitary Fermi gas. Within the framework of an extended $T$-matrix approximation, we calculate the single-particle density of states, as well as the single-particle spectral weight, to show that the so-called pseudogap phenomenon gradually disappears with increasing the magnitude of an effective magnetic field. In the highly spin-polarized regime, the calculated spin-polarization rate as a function of the effective magnetic field agrees well with the recent experiment on a $^6$Li Fermi gas. Although this experiment has been considered to be incompatible with the existence of the pseudogap in an {\it unpolarized} Fermi gas, our result clarifies that the observed spin-polarization rate in the highly spin-polarized regime and the pseudogap in the unpolarized limit can be explained in a consistent manner, when one correctly includes effects of population imbalance on single-particle excitations. Since it is a crucial issue to clarify whether the pseudogap exists or not in the BCS (Bardeen-Cooper-Schrieffer)-BEC (Bose-Einstein condensation) crossover regime of an ultracold Fermi gas, our results would be useful for the understanding of this strongly interacting fermion system.
\end{abstract}
\maketitle
%%%%%%%%%%%%%%%%%%%%%%%%%%%%%%%%%%%%%%%%%%%%%%%%%%%%%%%%%%%%%%%%%%%%%%%%%%%%%%
\par
\section{Introduction}
\par
In ultracold $^{40}$K and $^6$Li Fermi gases, one can experimentally tune the strength of a pairing interaction by adjusting the threshold energy of a Feshbach resonance\cite{Timmermans,Holland,Chin}. This unique property enables us to study the BCS-BEC crossover phenomenon\cite{Regal,Bartenstein,Zwierlein,Kinast}, where the character of a Fermi superfluid continuously changes from the weak-coupling BCS-type to the BEC of tightly bound molecules, as one increases the strength of a pairing interaction\cite{Eagles,Leggett,NSR,Randeria,SadeMelo,Ohashi,Strinati,Haussmann}. The intermediate coupling regime (which is also referred to as the BCS-BEC crossover region in the literature) is dominated by strong pairing fluctuations,  so that this regime is particularly useful for the study of various many-body effects in a systematic manner. The BCS-BEC crossover is one of the most exciting topics in cold Fermi gas physics\cite{Chin,Review}. 
\par
However, while the importance of pairing fluctuations in the BCS-BEC crossover regime of an ultracold Fermi gas is now widely accepted, details of resulting many-body phenomena are still in debate. In particular, the existence of the pseudogap phenomenon has been extensively discussed both theoretically\cite{Tsuchiya,Watanabe,Watanabe2,Chen2009,Hu2010,Bulgac} and experimentally\cite{Stewart,Gaebler,Perali,Kohl,Nascimbene2,Nascimbene,Navon}. While the photoemission-type experiments on a three-dimensional\cite{Stewart,Gaebler} and a two-dimensional\cite{Kohl} $^{40}$K Fermi gas support the pseudogap scenario\cite{Tsuchiya,Watanabe,Watanabe2,Chen2009,Hu2010,Bulgac}, the local pressure experiment\cite{Nascimbene2}, as well as the experiment on the spin polarization rate\cite{Nascimbene} on $^6$Li Fermi gases support the Fermi liquid theory. Between the latter two experiments, it has been pointed out that the former can be explained by a strong-coupling theory including pseudogap effects\cite{Watanabe}. Thus, on the viewpoint of the pseudogap scenario, it is an important challenge to explain the latter experiment on the spin-polarization rate within the framework of a pseudogap theory. Since the normal Fermi liquid is characterized by long-lived Fermi quasi-particles near the Fermi surface, it is quite different from the pseudogapped state, where the lifetime of Fermi atoms is very short near the Fermi level, because of the formation of preformed Cooper pairs. Thus, whether an ultracold Fermi gas in the BCS-BEC crossover region is a Fermi liquid or a pseudogapped Fermi gas is a crucial issue in cold Fermi gas physics. Since the pseudogap phenomenon has been extensively discussed in the under-doped regime of high-$T_{\rm c}$ cuprates\cite{Yanase,Pines,Kampf,Chakravarty}, this problem is also important to assess the preformed-pair scenario discussed in this strongly correlated electron system\cite{Yanase}. 
\par
In this paper, we investigate strong-coupling properties of a unitary Fermi gas with population imbalance, to clarify whether or not the recent experiment on the spin polarization rate\cite{Nascimbene} can be explained within the pseudogap scenario. In considering this problem, we note that Ref.\cite{Nascimbene} assumes that, if the pseudogap really exists in an unpolarized Fermi gas, this effect should still remain even in the spin-polarized case. Under this assumption, they extrapolate their experimental data in the highly spin-polarized regime to the unpolarized limit, leading to the conclusion that the pseudogap does not exist in the absence of population imbalance. Thus, to accomplish our purpose, a crucial key is to clarify to what extent the pseudogap phenomenon is sensitive/insensitive to the presence of population imbalance. 
\par
In the spin-polarized case, it is known that the ordinary Gaussian fluctuation theory developed by Nozi\`eres and Schmitt-Rink\cite{NSR}, as well as the (non-self-consistent) $T$-matrix approximation\cite{Strinati,Tsuchiya}, that have been extensively used to clarify various interesting BCS-BEC crossover phenomena in the unpolarized case, breakdown\cite{Liu,Parish}. To overcome this serious problem, we have recently presented a minimal extension of the $T$-matrix theory to include higher order pairing fluctuations\cite{Kashimura}. The calculated spin susceptibility in this extended $T$-matrix approximation (ETMA)\cite{Kashimura} quantitatively agrees well with the recent experiment on a $^6$Li Fermi gas\cite{Sanner}, without introducing any fitting parameter. In this paper, we also employ this strong-coupling theory, to examine the pseudogap phenomenon in a spin-polarized unitary Fermi gas. Within the same framework, we also calculate the spin polarization rate, to compare our result with the recent experiment on a $^6$Li unitary Fermi gas\cite{Nascimbene}. 
\par
This paper is organized as follows. In Sec. II, we explain the extended $T$-matrix approximation (ETMA). In Sec. III, we examine the pseudogap phenomenon in the presence of population imbalance. In Sec. IV, we discuss the spin-polarization rate as a function of an effective magnetic field, to compare our result with the recent experiment on a $^6$Li unitary Fermi gas\cite{Nascimbene}. Throughout this paper, we set $\hbar=k_{\text{B}}=1$, and the system volume $V$ is taken to be unity, for simplicity.
\par
%%%%%%%%%%%%%%%%%%%%%%%%%%%%%%%% figure %%%%%%%%%%%%%%%%%%%%%%%%%%%%%%%%%%%%%%%
\begin{figure}[t]   
\begin{center}
\includegraphics[keepaspectratio,scale=1]{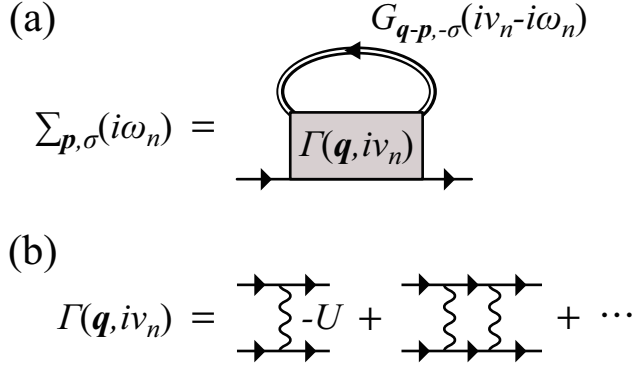}
\caption{(a) Self-energy correction $\Sigma_{\bm{p}, \sigma} (i\omega_n)$ in the extended $T$-matrix approximation (ETMA). (b) ETMA particle-particle vertex function $\Gamma (\bm{q}, i\nu_n)$. The solid line and the double solid line describe the bare Green's function $G^0_{\bm{p},\sigma} (i\omega_n)$ and the dressed Green's function $G_{\bm{p},\sigma} (i\omega_n)$, respectively. The wavy line represents the attractive interaction $-U (<0)$.
}
\label{fig1}
\end{center}    
\end{figure}
%%%%%%%%%%%%%%%%%%%%%%%%%%%%%%%%%%%%%%%%%%%%%%%%%%%%%%%%%%%%%%%%%%%%%%%%%%%%%%%
\par
\section{Formulation}
\par
We consider a two-component unitary Fermi gas with population imbalance, described by the BCS Hamiltonian,
\begin{equation}
H = \sum_{\bm{p}, \sigma} \xi_{\bm{p}, \sigma} c^\dagger_{\bm{p}, \sigma} c_{\bm{p}, \sigma}
- U  \sum_{\bm{p}, \bm{p}', \bm{q}} c^\dagger_{\bm{p}+\bm{q}/2, \up} c^\dagger_{-\bm{p}+\bm{q}/2, \dwn} c_{-\bm{p}'+\bm{q}/2, \dwn} c_{\bm{p}'+\bm{q}/2, \up}.
\label{H}
\end{equation}
Here, $c^\dagger_{\bm{p}, \sigma}$ is the creation operator of a Fermi atom with momentum $\bm{p}$ and pseudo-spin $\sigma (=\uparrow, \downarrow)$, describing two atomic hyperfine states. $\xi_{\bm{p}, \sigma} = \varepsilon_{\bm{p}}-\mu_\sigma =p^2/(2m)-\mu_\sigma$ is the kinetic energy, measured from the Fermi chemical potential $\mu_\sigma$ (where $m$ is an atomic mass). In the presence of population imbalance, one obtains $\mu_\uparrow\ne\mu_\downarrow$, so that atoms feel an effective magnetic field $h=[\mu_\uparrow-\mu_\downarrow]/2$. $-U$ is an assumed pairing interaction associated with a Feshbach resonance, which is related to the $s$-wave scattering length $a_s$ as,
\begin{equation}
\frac{4 \pi a_s}{m} = \frac{-U}{1-U\sum_{\bm{p}}^{\omega_{\rm c}} \frac{1}{2\varepsilon_{\bm{p}}}},
\label{REN}
\end{equation}
where $\omega_{\rm c}$ is a high-energy cutoff. In this paper, we consider a unitary Fermi gas, by setting $a_s^{-1}=0$. In addition, we assume a uniform Fermi gas, ignoring effects of a harmonic trap, for simplicity. 
\par
Strong-coupling effects on single-particle excitations are conveniently described by the self-energy correction $\Sigma_{\bm{p},\sigma}(i\omega_n)$ in the single-particle thermal Green's function, 
\begin{equation}
G_{\bm{p},\sigma} (i\omega_n) = \frac{1}{\left[ G^0_{\bm{p},\sigma} (i\omega_n) \right]^{-1}-\Sigma_{\bm{p}, \sigma}(i\omega_n)}.
\label{GF}
\end{equation}
Here, $\omega_n$ is the fermion Matsubara frequency, and $G^0_{\bm{p},\sigma} (i\omega_n) = [i \omega_n - \xi_{\bm{p}, \sigma}]^{-1}$ is the bare Green's function. In the extended $T$-matrix approximation (ETMA)\cite{Kashimura}, the self-energy $\Sigma_{\bm{p},\sigma}(i\omega_n)$ is diagrammatically described as Fig. \ref{fig1}, which gives
\begin{equation}
\Sigma_{\bm{p}, \sigma}(i\omega_n) = T \sum_{\bm{q}, i\nu_n} \Gamma (\bm{q},i\nu_n) G_{\bm{q}-\bm{p}, -\sigma} (i\nu_n-i\omega_n),
\label{SE}
\end{equation}
where $\nu_n$ is the boson Matsubara frequency. In Eq. (\ref{SE}), $-\sigma$ means the opposite component to the $\sigma$-component. The ETMA particle-particle vertex function $\Gamma (\bm{q},i\nu_n)$ has the form,
\begin{equation} \begin{split}
\Gamma (\bm{q}, i\nu_n) &= \frac{-U}{1-U \Pi (\bm{q}, i\nu_n)},
\label{VF}
\end{split} \end{equation}
where 
\begin{eqnarray}
\Pi (\bm{q}, i\nu_n) &=& T \sum_{\bm{p},i \omega_n} G^0_{\bm{p}+\bm{q}/2,\up} (i\nu_n+i\omega_n) G^0_{-\bm{p}+\bm{q}/2,\dwn} (-i\omega_n) 
\nonumber\\
&=& - \sum_{\bm{p}} \frac{1-f(\xi_{\bm{p}+\bm{q}/2, \up}) - f(\xi_{-\bm{p}+\bm{q}/2, \dwn})}{i \nu_n-\xi_{\bm{p}+\bm{q}/2, \up} - \xi_{-\bm{p}+\bm{q}/2, \dwn}}
\label{TC}
\end{eqnarray}
is the lowest order pair correlation function, describing pairing fluctuations. 
\par
We briefly note that the self-energy in the ordinary $T$-matrix approximation (TMA) \cite{Strinati,Tsuchiya,Watanabe,Watanabe2} is also given by Eq. (\ref{SE}) where the dressed Green's function $G$ is simply replaced by the bare one $G^0$ (which diagrammatically corresponds to replace the double solid line ($G$) in Fig. \ref{fig1}(a) by the solid line ($G^0$)). Because of this difference, while the TMA susceptibility unphysically becomes negative in the crossover region, the ETMA correctly gives positive susceptibility in the whole BCS-BEC crossover region. To explain the reason for this improvement in a simple manner, it is convenient to replace the particle-particle vertex function $\Gamma({\bm q},i\nu_n)$ in Eq. (\ref{VF}) by the bare interaction $-U$. In this simple case, the ETMA spin susceptibility $\chi$, which has the vertex correction being consistent with the ETMA self-energy in Eq. (\ref{SE}), involves the RPA (random phase approximation)-type infinite series of bubble diagrams, as $\chi\sim \chi_0/(1+U\chi_0)$ (where $\chi_0$ is the lowest order spin susceptibility). On the other hand, this RPA series is truncated at $O(U)$ as $\chi\sim\chi_0(1-U\chi_0)$ in the TMA. As a result, while the ETMA susceptibility is simply suppressed with increasing the interaction strength $U$, the TMA susceptibility unphysically becomes negative when $U\chi_0<0$. For more details, we refer to Ref. \cite{Kashimura}.
\par
%%%%%%%%%%%%%%%%%%%%%%%%%%%%%% figure %%%%%%%%%%%%%%%%%%%%%%%%%%%%%%%%%%%%%%%%
\begin{figure}[t]   
\begin{center}
\includegraphics[keepaspectratio,scale=0.5]{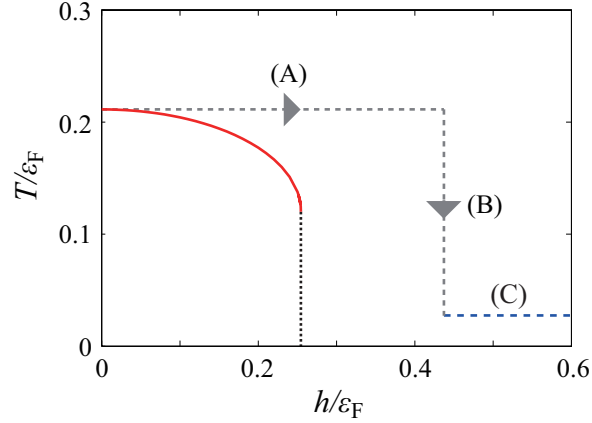}
\caption{(Color online) We examine effects of population imbalance on single-particle excitations in a unitary Fermi gas along the path (A)-(B). In the case of  path (A), we fix the temperature as $T=T_{\rm c}(h=0)$. In the case of path (B), the effective magnetic field is fixed as $h/\varepsilon_{\text{F}}=0.43$. In the region (C) ($T/\varepsilon_{\text{F}}=0.03$, and $h/\varepsilon_{\rm F}\ge 0.43$), we calculate the spin polarization rate $P$, as a function of the effective magnetic field $h$, where Ref. \cite{Nascimbene} measured this quantity in a $^6$Li Fermi gas. The solid line shows the calculated $T_{\rm c}$ in the unitary limit. $\varepsilon_{\rm F}=k_{\rm F}^2/(2m)$ is the Fermi energy in an unpolarized free Fermi gas, where $k_{\rm F}=(3\pi^2 N)^{1/3}$ is the Fermi momentum.
}
\label{fig2}
\end{center}    
\end{figure}
%%%%%%%%%%%%%%%%%%%%%%%%%%%%%%%%%%%%%%%%%%%%%%%%%%%%%%%%%%%%%%%%%%%%%%%%%%%%%%
\par
Here, we explain our strategy to attack the pseudogap problem mentioned in Sec. I. We first examine effects of population imbalance on the pseudogap phenomenon in a unitary Fermi gas ($a_s^{-1}=0$) along the path (A)-(B) in Fig. \ref{fig2}. For this purpose, we calculate the single-particle spectral weight,
\begin{equation}
A_{\bm{p},\sigma} (\omega) =-
\frac{1}{\pi} \text{Im} \left[ G_{\bm{p}, \sigma} (i \omega_n \rightarrow \omega + i\delta) \right],
\label{SW}
\end{equation}
as well as the single-particle density of states,
\begin{equation}
\rho_{\sigma} (\omega) = \sum_{\bm{p}} A_{\bm{p},\sigma} (\omega), 
\label{DOS}
\end{equation}
where $G_{\bm{p}, \sigma} (i \omega_n \rightarrow \omega + i\delta)$ is the analytic continued ETMA dressed Green's function (where $\delta$ is an infinitesimally small positive number). In calculating these quantities, the Fermi chemical potential $\mu_\sigma$ in the Green's function is determined from the equation for the number $N_\sigma$ of Fermi atoms in the $\sigma$-component, given by
\begin{equation}
N_\sigma = T \sum_{\bm{p},i\omega_n} G_{\bm{p},\sigma} (i\omega_n) e^{i\delta\omega_n}.
\label{Num}
\end{equation}
The total particle number $N=N_\uparrow+N_\downarrow$ is fixed, in determining $\mu_\sigma$ along the path (A)-(B). 
\par
We then consider the region (C) in Fig. \ref{fig2}, where the spin-polarization rate,
\begin{equation}
P={N_{\uparrow}-N_\downarrow \over N_\uparrow+N_\downarrow},
\label{polari}
\end{equation}
has been recently measured\cite{Nascimbene}. In this regime, we calculate this quantity, to see to what extent the ETMA (which gives the pseudogap in the unpolarized case) can explain this experiment\cite{Nascimbene} (which has been considered to be incompatible with the existence of the pseudogap in the unpolarized case).
\par
As shown in Fig. \ref{fig2}, both the path (A)-(B) and the region (C) are in the normal state. In this figure, the superfluid phase transition temperature $T_{\rm c}$ is determined from the $T_{\rm c}$-equation\cite{Kashimura,note2,Thouless},
\begin{eqnarray}
0=\sum_{\bm{p}} \left[ 
\frac{\tanh{\left( \frac{\xi_{\bm{p}, \up}}{2T_{\text{c}}} \right)}  + \tanh{\left( \frac{\xi_{\bm{p}, \dwn}}{2T_{\text{c}}} \right)}}
{\xi_{\bm{p},\uparrow}+\xi_{\bm{p},\downarrow}}
-\frac{1}{\varepsilon_{\bm{p}}}
\right].
\label{ThoulessC}
\end{eqnarray}
In Fig. \ref{fig2}, $T_{\rm c}$ vanishes at $h\simeq 0.255\varepsilon_{\rm F}~(\equiv h_{\rm c})$, as expected from the analogy to metallic superconductivity under an external magnetic field (where $\varepsilon_{\rm F}$ is the Fermi energy in an unpolarized free Fermi gas). In this regard, we note that the phase-separated region which consists of the superfluid phase and the normal phase would actually exist in a real ultracold Fermi gas around the region $h\gesim h_{\rm c}$\cite{Shin}, which, however, cannot be described by the present theory assuming a uniform gas. In this regard, we note that the phase transition temperature in this regime would be lower than $T_{\rm c}(h=0)$ in the unpolarized limit\cite{Shin}. In addition, it has been experimentally confirmed that the region (C) in Fig. \ref{fig2} is in the normal state\cite{Nascimbene,Shin,note}. Thus, the system would be still in the normal state along the path (A)-(B), as well as in the region (C), even in a more sophisticated theory which can treat the phase separation.
\par
%%%%%%%%%%%%%%%%%%%%%%%%%%%%%%%%%%%%%%%%%%%%%%%%%%%%%%%%%%%%%%%%%%%%%%%%%%%%%%
\begin{figure}[t]   
\begin{center}
\includegraphics[keepaspectratio,scale=0.5]{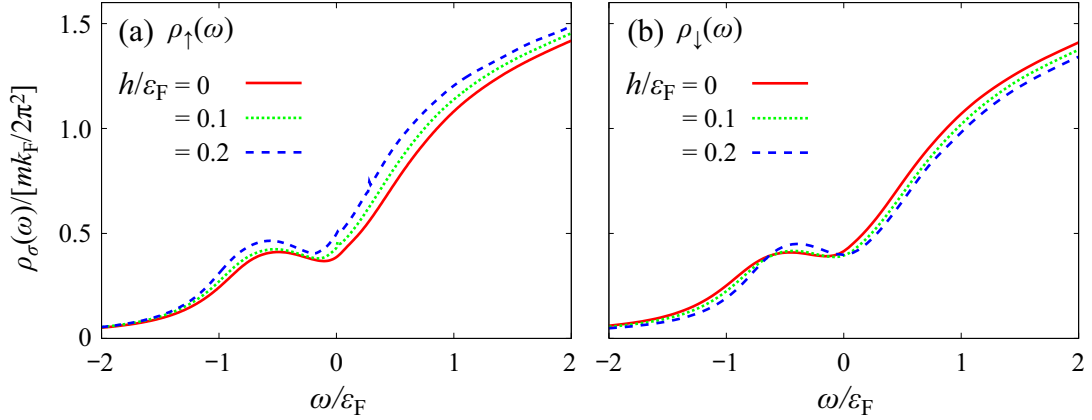}
\caption{
(Color online) Calculated single-particle density of states $\rho_\sigma(\omega)$ at $T_{\text{c}}$ in a unitary Fermi gas with population imbalance. 
}
\label{fig3}
\end{center}    
\end{figure}
%%%%%%%%%%%%%%%%%%%%%%%%%%%%%%%%%%%%%%%%%%%%%%%%%%%%%%%%%%%%%%%%%%%%%%%%%%%%%%
\par
\section{Single-particle excitations and pseudogap phenomenon in a spin-polarized unitary Fermi gas}
\par
Figure \ref{fig3} shows the single-particle density of states $\rho_\sigma(\omega)$ at $T_{\rm c}$ in the unitarity limit. In the absence of the population imbalance ($h=0$), one sees a dip structure around $\omega=0$. Since the superfluid order parameter vanishes at $T_{\rm c}$, this is just the pseudogap originating from strong pairing fluctuations. This dip structure exists even in the presence of population imbalance, when one moves along the $T_{\rm c}$-line given in Fig. \ref{fig2}.
\par
%%%%%%%%%%%%%%%%%%%%%%%%%%%%%%%%%%%%%%%%%%%%%%%%%%%%%%%%%%%%%%%%%%%%%%%%%%%%%%
\begin{figure}[t]   
\begin{center}
\includegraphics[keepaspectratio,scale=0.5]{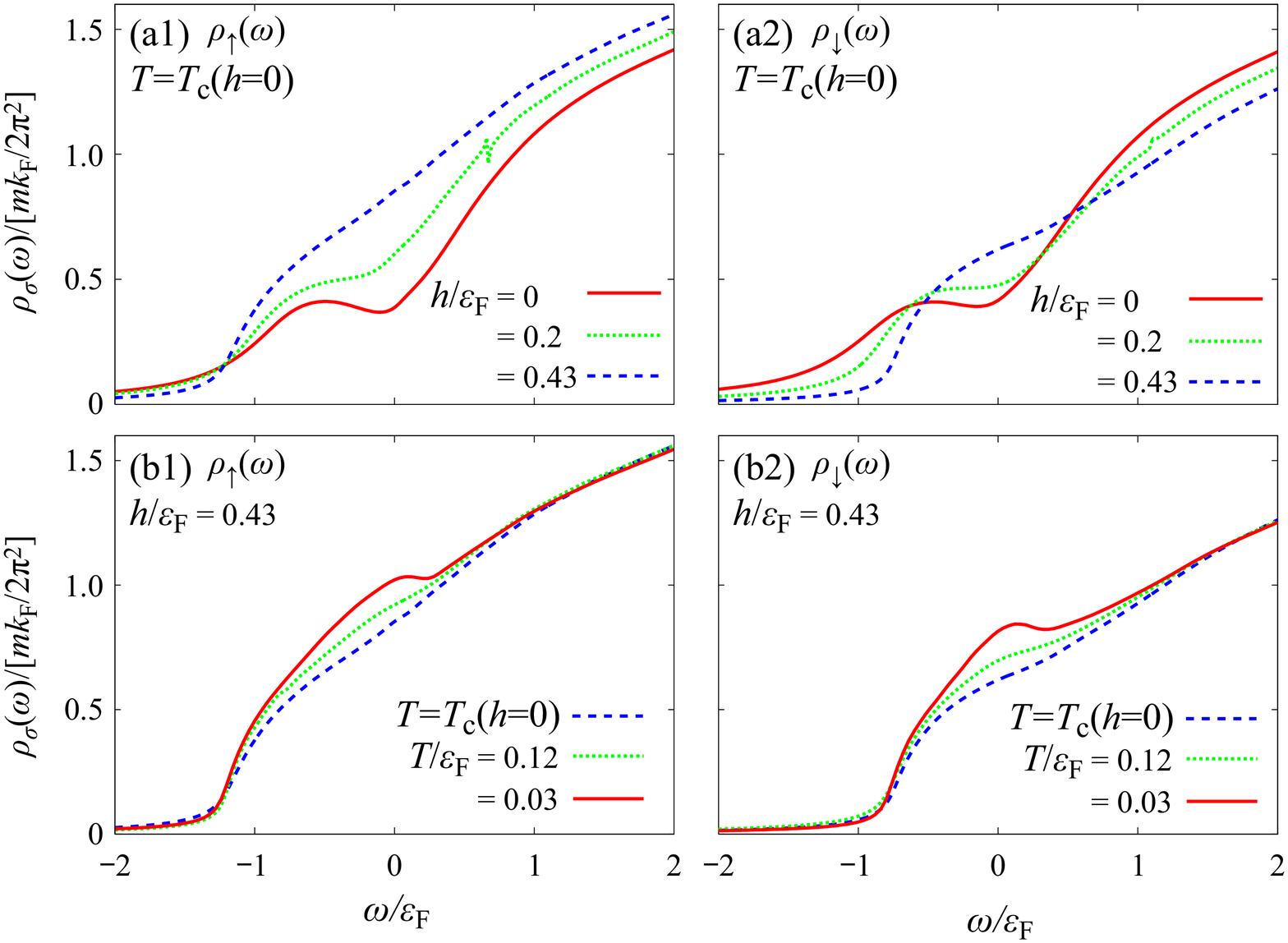}
\caption{
(Color online) Calculated single-particle density of states $\rho_\sigma(\omega)$ along the path (A) (upper panels) and along the path (B) (lower panels). In the upper and lower panels, we set $T=T_{\rm c}(h=0)$ and $h/\varepsilon_{\rm F}=0.43$, respectively. 
}
\label{fig4}
\end{center}    
\end{figure}
%%%%%%%%%%%%%%%%%%%%%%%%%%%%%%%%%%%%%%%%%%%%%%%%%%%%%%%%%%%%%%%%%%%%%%%%%%%%%%
\par
When we move along the path (A) in Fig. \ref{fig2}, since we gradually go away from the superfluid region, the pseudogap in $\rho_\sigma(\omega)$ gradually disappears, as shown in Figs. \ref{fig4}(a1) and (a2). At the end of the path (A) ($h/\varepsilon_{\rm F}=0.43$), apart from details, the overall structure of the density of states $\rho_\sigma(\omega)$ is close to the density of states $\rho_\sigma^0(\omega)$ in a free Fermi gas, given by
\begin{equation}
\rho^0_\sigma(\omega)={mk_{\rm F} \over 2\pi^2}\sqrt{\omega+\mu_\sigma}~~(\omega\ge-\mu_\sigma),
\label{dos0}
\end{equation}
where $k_{\rm F}$ is the Fermi momentum in an unpolarized free Fermi gas. However, when we further move along the path (B), a dip structure revives, not around $\omega=0$, but above $\omega=0$, as shown in Figs. \ref{fig4}(b1) and (b2). 
\par
%%%%%%%%%%%%%%%%%%%%%%%%%%%%%%%%%%%%%%%%%%%%%%%%%%%%%%%%%%%%%%%%%%%%%%%%%%%%%%
\begin{figure}[t]   
\begin{center}
\includegraphics[keepaspectratio,scale=0.7]{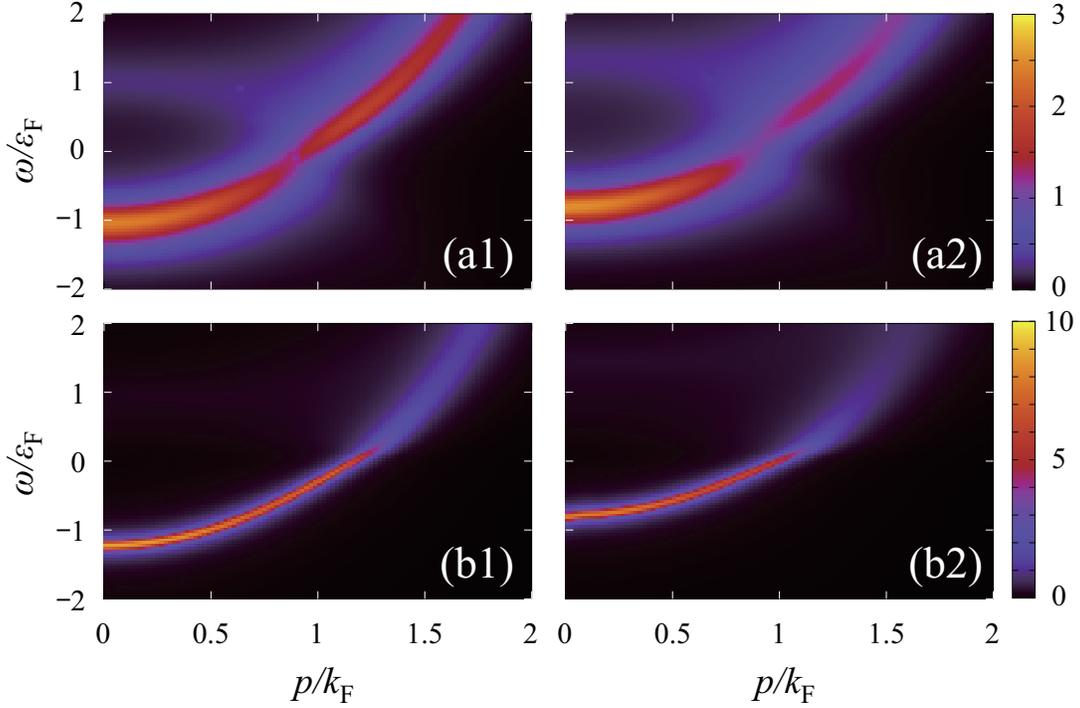}
\caption{(Color online) Calculated intensity of the single-particle spectral weight $A_{{\bm p},\sigma}(\omega)$, normalized by $\varepsilon_{\rm F}^{-1}$. The upper two panels show the results at $T_{\rm c}$ when $h/\epsilon_{\text{F}}=0.2$. The lower two panels show the results at the end of the path (B) ($h/\epsilon_{\text{F}}=0.43$, and $T/\epsilon_{\text{F}} =0.03$). The left (right) two panels show $A_{{\bm p},\uparrow}(\omega)$ ($A_{{\bm p},\downarrow}(\omega)$). 
}
\label{fig5}
\end{center}    
\end{figure}
%%%%%%%%%%%%%%%%%%%%%%%%%%%%%%%%%%%%%%%%%%%%%%%%%%%%%%%%%%%%%%%%%%%%%%%%%%%%%%
\par
We point out that the origin of the dip structure at the end of the path (B) is different from the ordinary pseudogap mechanism working in the case of Fig. \ref{fig3}. In the latter case, the pseudogap may be viewed as a particle-hole coupling phenomenon induced by strong-pairing fluctuations\cite{Tsuchiya}. Indeed, as shown in Ref. \cite{Hanai}, under the assumption that pairing fluctuations are strong in the low-energy and low-momentum region near $T_{\rm c}$, when we approximate the ETMA self-energy in Eq. (\ref{SE}) to $\Sigma_{{\bm p},\sigma}(i\omega_n)\simeq G_{-\bm{p}, -\sigma} (-i\omega_n)\times T \sum_{\bm{q}, i\nu_n} \Gamma (\bm{q},i\nu_n)$, we obtain
\begin{equation}
\Sigma_{{\bm p},\sigma}(i\omega_n)
\simeq
{{\tilde \Delta}_{{\rm pg},\sigma}^2(i\omega) \over i\omega+\xi_{{\bm p},-\sigma}}.
\end{equation}
Here, 
\begin{eqnarray}
{\tilde \Delta}_{{\rm pg},\sigma}^2(i\omega_n)=
{2\Delta_{\rm pg}^2 
\over 
1+\sqrt{1-
{\displaystyle 2\Delta^2_{\rm pg} 
\over \displaystyle
(i\omega_n-\xi_{{\bm p},\sigma})(i\omega_n+\xi_{{\bm p},-\sigma})
}}}
\label{eq.ETMAapp}
\end{eqnarray}
is the ETMA pseudogap parameter\cite{Hanai}, where $\Delta^2_{\rm pg}=-T \sum_{\bm{q}, i\nu_n} \Gamma (\bm{q},i\nu_n)$\cite{Tsuchiya,Chen2009}. Substituting Eq. (\ref{eq.ETMAapp}) into Eq. (\ref{GF}), we obtain
\begin{equation}
G_{{\bm p},\sigma}(i\omega_n)=
{1 \over \displaystyle i\omega_n-\xi_{{\bm p},\sigma}-
{{\tilde \Delta}_{{\rm pg},\sigma}^2(i\omega_n) \over i\omega_n+\xi_{{\bm p},-\sigma}}},
\label{eq.GGG}
\end{equation}
which physically means that pairing fluctuations described by the ETMA pseudogap parameter ${\tilde \Delta}_{{\rm pg},\sigma}$ couple the particle branch $\omega=\xi_{{\bm p},\sigma}$ with the hole branch $\omega=-\xi_{{\bm p},-\sigma}$. Indeed, in each of Figs. \ref{fig5}(a1) and (a2), we slightly see a broad spectral peak of the hole branch, in addition to the particle branch. In this figure, this particle-hole coupling is found to modify the single-particle excitation spectrum around $\omega=0$, leading to the pseudogap structure in Fig. \ref{fig3}.
\par
In contrast to the pseudogapped case shown in Figs. \ref{fig5}(a1) and (a2), such particle-hole coupling is not seen in the spectral weight $A_{{\bm p},\sigma}(\omega)$ at the end of path (B), as shown in Figs. \ref{fig5}(b1) and (b2). In each of these panels, we see a sharp spectral peak line describing particle dispersion below $\omega=0$, which becomes broad when $\omega\gesim 0$. Since the width of the spectral peak is directly related to the quasi-particle lifetime $\tau$, the sharp spectral peak in the negative energy region ($\omega\lesssim 0$) indicates that the Pauli blocking suppresses quasi-particle scatterings there. On the other hand, the broad spectral peak in the positive energy region ($\omega\gesim 0$) indicates frequent quasi-particle scatterings, leading to short quasi-particle lifetime. Since the density of states $\rho_\sigma(\omega)$ is given by the summation of the spectral weight $A_{{\bm p},\sigma}(\omega)$ with respect the momentum ${\bm p}$ (See Eq. (\ref{DOS}).), the broad spectral weight above $\omega=0$ is considered to decrease the magnitude of the density of states $\rho_\sigma(\omega)$, leading to the dip structure when $h/\varepsilon_{\rm F}=0.43$ and $T/\varepsilon_{\rm F}=0.03$\cite{notez}.
\par
%%%%%%%%%%%%%%%%%%%%%%%%%%%%%%%%%%%%%%%%%%%%%%%%%%%%%%%%%%%%%%%%%%%%%%%%%%%%%%
\begin{figure}[t]
\begin{center}
\includegraphics[keepaspectratio,scale=0.5]{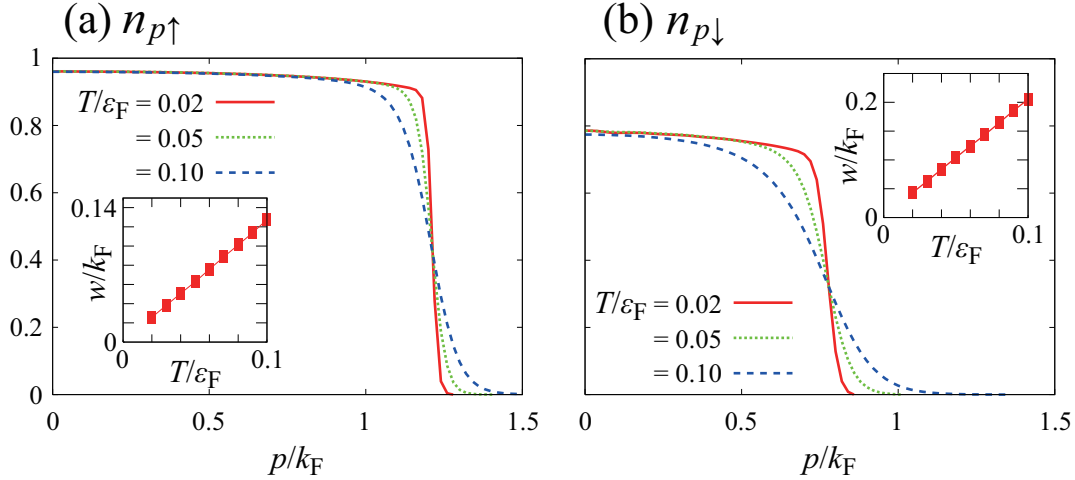}
\caption{
(Color online) Calculated momentum distribution $n_{{\bm p},\sigma}=\langle c_{{\bm p},\sigma}^\dagger c_{{\bm p},\sigma}\rangle$ in a highly spin-polarized unitary Fermi gas ($h/\epsilon_{\text{F}}=0.8$). (a) $n_{{\bm p},\uparrow}$. (b) $n_{{\bm p},\downarrow}$. In each panel, the inset shows the momentum width $w=p_2-p_1$ of the Fermi surface edge, where $p_1$ ($p_2$) is the momentum satisfying $n_{{p_1}\uparrow}=0.2~(n_{{p_2}\downarrow}=0.75$) in the case of the $\uparrow$-spin component, and satisfying $n_{{p_1}\uparrow}=0.2~(n_{{p_2}\downarrow}=0.6$) in the case of the $\downarrow$-spin component. 
}
\label{fig6}
\end{center}    
\end{figure}
%%%%%%%%%%%%%%%%%%%%%%%%%%%%%%%%%%%%%%%%%%%%%%%%%%%%%%%%%%%%%%%%%%%%%%%%%%%%%%
\par
The above discussion at the end of path (B) makes us expect the existence of clear Fermi surfaces in the low temperature region of a highly spin-polarized Fermi gas, even in the unitarity limit. To confirm this, we show in Fig. \ref{fig6} the momentum distribution,
\begin{equation}
n_{{\bm p},\sigma}=\langle c_{{\bm p},\sigma}^\dagger c_{{\bm p},\sigma}\rangle
=T\sum_{i\omega_n}G_{{\bm p},\sigma}(i\omega)e^{i\delta\omega_n},
\end{equation}
in a highly spin-polarized Fermi gas ($h/\varepsilon_{\rm F}=0.8$). As expected, $n_{{\bm p},\sigma}$ exhibits a clear ``Fermi surface edge", which becomes sharper at lower temperatures. Although we cannot calculate this quantity at $T=0$ because of computational problems, from the extrapolation of our results shown in the insets in Fig. \ref{fig6}, the momentum distribution $n_{{\bm p},\sigma}$ at $T=0$ is expected to be very close to the step function, as in the case of a free Fermi gas at $T=0$.
\par
The existence of the sharp Fermi surface edge indicates the validity of the Fermi quasi-particle picture in this regime, so that the pseudogapped Green's function in Eq. (\ref{eq.GGG}) would be no longer valid there. In the quasi-particle picture, expanding the real part of the analytic continued self-energy ${\rm Re}[\Sigma_{{\bm p},\sigma}(i\omega_n\to\omega+i\delta)]$ to $O(\omega)$, we obtain the (analytic continued) quasi-particle Green's function as,
\begin{equation}
G_{{\bm p},\sigma}(\omega+i\delta)=
{Z_{{\bm p},\sigma} \over \omega-{\tilde \xi}_{{\bm p},\sigma}-{\rm Im}{\tilde \Sigma}_{{\bm p},\sigma}(\omega+i\delta)}.
\label{quasi}
\end{equation}
Here, 
\begin{equation}
Z_{{\bm p},\sigma}=
\left[
1-{\partial {\rm Re}[\Sigma_{{\bm p},\sigma}(\omega+i\delta)] \over \partial \omega}
\right]^{-1}
\end{equation}
is a renormalization factor in the $\sigma$-spin component, and ${\tilde \Sigma}_{{\bm p},\sigma}(\omega+i\delta)=Z_{{\bm p},\sigma}\Sigma_{{\bm p},\sigma}(\omega+i\delta)$. ${\tilde \xi}_{{\bm p},\sigma}=p^2/(2{\tilde m})-{\tilde \mu}_{{\bm p},\sigma}$ is the (renormalized) kinetic energy of a Fermi quasi-particle, where ${\tilde m}=m/Z_{{\bm p},\sigma}$ and ${\tilde \mu}_{{\bm p},\sigma}=Z_{{\bm p},\sigma}[\mu_\sigma-\Sigma_{{\bm p},\sigma}(0)]$. Strictly speaking, the quasi-particle Green's function in Eq. (\ref{quasi}) is only valid for the low-energy region. However, when we simply use this expression in calculating the momentum distribution $n_{{\bm p},\sigma}$, ignoring lifetime effects described by ${\rm Im}{\tilde \Sigma}_{{\bm p},\sigma}(\omega+i\delta)$ for simplicity, we obtain, at $T=0$,
\begin{equation}
n_{{\bm p},\sigma}=Z_{{\bm p},\sigma}\Theta(-{\tilde \xi}_{{\bm p},\sigma}),
\label{eq.quasi1}
\end{equation}
where $\Theta(x)$ is the step function. From the comparison of Eq. (\ref{eq.quasi1}) with Fig. \ref{fig6}, the renormalization factors are roughly evaluated as $Z_{{\bm p}\uparrow}\simeq 0.95$ and $Z_{{\bm p},\downarrow}\simeq 0.8$.
\par
We note that, although the Fermi quasi-particle picture is valid for a highly spin-polarized unitary Fermi gas, it does not mean the validity of this description in the unpolarized case. As shown in Fig. \ref{fig3}, the pseudogap actually exists in the absence of population imbalance. Although Ref. \cite{Nascimbene} assumes that the shape of the density of states remains unchanged in the presence of a finite magnetic field $h$, the ETMA clearly shows that the character of a unitary Fermi gas actually continuously changes from the pseudogapped gas to a gas of Fermi quasi-particles, along the path (A)-(B) given in Fig. \ref{fig2}. 
\par
%%%%%%%%%%%%%%%%%%%%%%%%%%%%%%%%%%%%%%%%%%%%%%%%%%%%%%%%%%%%%%%%%%%%%%%%%%%%%%
\begin{figure}[t]   
\begin{center}
\includegraphics[keepaspectratio,scale=0.5]{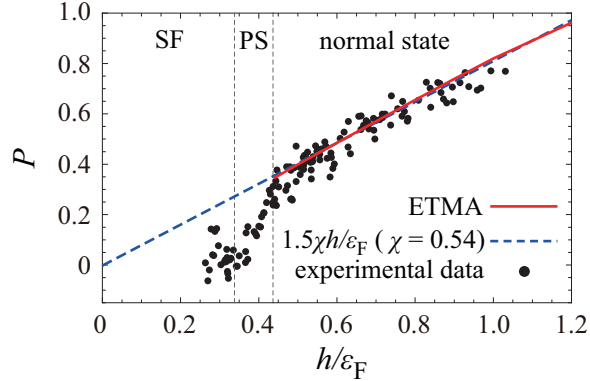}
\caption{(Color online) Spin polarization rate $P$, as a function of the effective magnetic field $h$ in a unitary Fermi gas. The sold line shows the ETMA result calculated in the region (C) in Fig. \ref{fig2}. Filled circles are experimental data on a $^6$Li unitary Fermi gas\cite{Nascimbene}. We also show the fitting line given in Ref. \cite{Nascimbene} (dashed line). In this figure, `PS' and `SF' mean the phase separated region and the superfluid phase, respectively\cite{note}.
}
\label{fig7}
\end{center}    
\end{figure}
%%%%%%%%%%%%%%%%%%%%%%%%%%%%%%%%%%%%%%%%%%%%%%%%%%%%%%%%%%%%%%%%%%%%%%%%%%%%%%
\par
\section{Spin polarization rate in a highly spin-polarized unitary Fermi gas}
\par
Figure \ref{fig7} shows the spin polarization rate $P$, as a function of the effective magnetic field $h$ in a unitary Fermi gas. The calculated spin polarization rate in the region (C) agrees well with the recent experiment on a $^6$Li Fermi gas\cite{Nascimbene}. We emphasize that there is no fitting parameter in obtaining our result, which indicates the validity of the ETMA for the study of a highly spin-polarized Fermi gas.
\par
So far, the experiment shown in Fig. \ref{fig7} has been considered to deny the existence of the pseudogap in an unpolarized unitary Fermi gas\cite{Nascimbene}, because of the fact that the fitting line obtained in the normal state ($h/\varepsilon_{\rm F}\gesim 0.43$) crosses the $y$-axis ($h=0$) at $P=0$, as shown in Fig. \ref{fig7}\cite{note4}. However, although our result also gives the same result when it is extrapolated to $h=0$, the pseudogap actually exists in the unpolarized limit, as shown in Fig. \ref{fig3}. That is, the spin-polarization rate $P$ observed in a highly spin-polarized $^6$Li Fermi gas does not contradict with the existence of the pseudogap in an unpolarized Fermi gas.
\par
%%%%%%%%%%%%%%%%%%%%%%%%%%%%%%%%%%%%%%%%%%%%%%%%%%%%%%%%%%%%%%%%%%%%%%%%%%%%%%%
\section{summary} 
\par
To summarize, we have discussed single-particle excitations in the normal state of a unitary Fermi gas with population imbalance. Within the framework of an extended $T$-matrix approximation, we have calculated the single-particle density of states, as well as the single-particle spectral weight, to clarify how the presence of population imbalance affects the pseudogap phenomenon. Within the same framework, we have also calculated the spin polarization rate as a function of an effective magnetic field.
\par
Along the path (A)-(B) in Fig. \ref{fig2}, we showed that the character of a unitary Fermi gas continuously changes from the pseudogapped Fermi gas to a gas of Fermi quasi-particles. That is, the dip structure in the density of states around $\omega=0$ originating from strong pairing fluctuations gradually disappears, when one moves along this path. In the highly spin polarized regime, a clear Fermi surface edge is obtained in each spin component ($\sigma=\uparrow,\downarrow$). From the extrapolation of our result down to $T=0$, the momentum distribution in this regime looks having discontinuity at the Fermi surface at $T=0$, as in the case of a normal Fermi liquid\cite{AGD}. Since the concept of Fermi liquid was originally introduced to describe {\it repulsively} interacting fermion systems, we need further study to check whether or not this idea can be also applicable to an {\it attractively} interacting Fermi gas. In this regard, at least, the existence of the discontinuity in the momentum distribution satisfies the condition for the Fermi liquid\cite{AGD}. Apart from this terminology, our results indicate that a spin-polarized Fermi gas is a useful system to examine how the pseudogapped Fermi gas changes into a quasi-particle Fermi gas in a systematic manner.
\par
In a highly spin-polarized regime (region (C) in Fig. \ref{fig2}), the calculated spin polarization rate $P$ as a function of an effective magnetic field $h$ agrees well with the recent experiment on a $^6$Li Fermi gas\cite{Nascimbene}, without introducing any fitting parameter. Although this experiment has been considered to contradict with the existence of the pseudogap in unpolarized Fermi gases, our result clarifies that this problem can be resolved, when one correctly includes the dependence of the density of states on the spin polarization. In addition, this agreement also indicates the validity of the extended $T$-matrix approximation (ETMA)\cite{Kashimura} for a spin-polarized Fermi gas.
\par
Although the present ETMA can describe the continuous change from a pseudogapped Fermi gas to a quasi-particle Fermi gas by adjusting the magnitude of an effective magnetic field $h$, there still exists room for improvement. To construct the phase diagram of a spin-polarized Fermi gas, we need to further extend this theory so that we can treat the phase separated region. In addition, the ETMA is not a fully self-consistent theory in the sense that it still uses the bare Green's function in the particle-particle vertex function $\Gamma({\bm q},i\nu_n)$. The improvement of this is also a crucial theoretical issue. In particular, details of the particle-particle vertex function are important in considering the so-called spin-polaron discussed in an extremely spin-polarized Fermi gas\cite{Stoof}. Thus, this improvement would enable us to treat the pseudogap phenomenon, quasi-particle Fermi gas, and spin polarons, in a unified manner. Since ultracold Fermi gases with population imbalance involve various interesting topics, our results would contribute to the further development of this active research field, in addition to the understanding of the pseudogap physics discussed in unpolarized Fermi gases. 
\par
%%%%%%%%%%%%%%%%%%%%%%%%%%%%%%%%%%%%%%%%%%%%%%%%%%%%%%%%%%%%%%%%%%%%%%%%%%%%%%%
\section*{Acknowledgements}
\par
We would like to thank Y. Endo, D. Inotani, and R. Hanai for useful discussions. Y. O. was supported by Grant-in-Aid for Scientific research from MEXT in Japan (No.25400418, No.25105511, No.23500056).
\par
%%%%%%%%%%%%%%%%%%%%%%%%%%%%%%%%%%%%%%%%%%%%%%%%%%%%%%%%%%%%%%%%%%%%%%%%%%%%%%%
\par

%%%%%%%%%%%%%%%%%%%%%%%%%%%%%%%%%%%%%%%%%%%%%%%%%%%%%%%%%%%%%%%%%%%%%%%%%%%%%%%

\begin{thebibliography}{99}
\bibitem{note2} As usual, the $T_{\rm c}$-equation (\ref{ThoulessC}) is obtained from the Thouless criterion\cite{Thouless}, stating that the superfluid phase transition occurs when the particle-particle vertex function $\Gamma({\bm q},i\nu_n)$ in Eq. (\ref{VF}) has a pole at ${\bm q}=0$ and $\nu_n=0$. In Eq. (\ref{ThoulessC}), we have taken $a_s^{-1}=0$.
\bibitem{note} In Ref. \cite{Nascimbene}, the phase transition from the superfluid phase to the normal state is reported to occur at $h/\epsilon_{\text{F}} \simeq 0.4$. In this regard, we note that Ref. \cite{Nascimbene} evaluates the spin polarization rate $P$ from experimental data under the condition of a fixed value of the total number $N$ of Fermi atoms. In this case, as shown in Ref. \cite{Shin}, with increasing the magnitude of the effective magnetic field $h$, the first order phase transition from the superfluid state to the phase separation should first occur. Then, one reaches the transition from phase-separated phase to the normal state, when the magnitude of $h$ is further increased. Including this, we have estimated from the experimental data\cite{Nascimbene} that the transition from the superfluid state to the phase separation occurs at $h/\epsilon_{\text{F}}=0.35$, and the transition from the the phase separated state to the normal state occurs at $h/\epsilon_{\text{F}}=0.43$. These boundaries are shown in Fig. \ref{fig7}.
\bibitem{notez} We briefly note that the particle dispersions in Figs. \ref{fig5} (b1) and (b2) are slightly flatter around $\omega=0$ than the quadratic dispersion of a free particle, which indicates the mass enhancement of Fermi quasi-particles there. As known in the ordinary Fermi liquid theory, this effect enhances the density of states around $\omega=0$, which is also considered to give the dip structures above $\omega=0$ in Figs. \ref{fig4}(b1) and (b2).
\bibitem{note4} Reference \cite{Nascimbene} argues that, if the pseudogap exists, the fitting line should cross the $y$-axis at a negative value of $P$.
\end{thebibliography}
\end{document}